\documentclass[runningheads]{llncs}

\usepackage[T1]{fontenc}

\usepackage{graphicx}
\usepackage{amssymb}
\usepackage{amsmath}
\usepackage{booktabs}   
\usepackage{multirow}   
\usepackage{adjustbox}  

\begin{document}

\title{MAF: Multimodal Adaptive Few-shot Prompting for Sentiment Analysis with MLLMs}

\titlerunning{MAF}

\author{Hangling Xie\inst{1}\orcidID{0009-0002-0287-3211}} 

\institute{Nanjing University of Posts and Telecommunications, Wenyuan Road No.9, Xixia District, Nanjing City, Jiangsu Province, China\\
\email{xiehangling0@163.com}}

\maketitle

\begin{abstract}

Multimodal large language models (MLLMs) have demonstrated remarkable capabilities in understanding complex multimodal content. However, their performance in sentiment analysis exhibits acute sensitivity to prompt design, rendering static, uniformly applied prompts inherently suboptimal for capturing the nuanced multimodal cues that vary across inputs. To address this limitation, we propose a \textbf{M}ultimodal \textbf{A}daptive \textbf{F}ew-Shot Prompting (\textbf{MAF}) framework, which dynamically retrieves and integrates query-relevant demonstrations to elicit the sentiment reasoning capabilities of MLLMs in a context-sensitive manner. MAF constructs a demonstration retrieval module that holistically encodes facial expressions, scene context, and textual semantics, with a lip movement amplitude detection mechanism introduced for accurate speaker identification in multi-person scenarios. Departing from conventional fixed-weight fusion, a lightweight coefficient generation network is trained to output query-conditioned fusion weights in real time, enabling weighted aggregation of multimodal similarity scores to retrieve the top-\(K\) most informative demonstrations. Prediction stability is further enhanced through majority voting over multiple candidate outputs generated by the MLLM. Extensive experiments on public benchmark datasets demonstrate that MAF achieves substantial and consistent performance improvements over the corresponding backbone variants and remains competitive with strong multimodal sentiment-analysis baselines.

\end{abstract}
\section{Introduction}

Sentiment analysis aims to automatically identify emotions in text, speech, or video. It encompasses Multimodal Sentiment Analysis (MSA)~\cite{zadeh2018} for single clips and Emotion Recognition in Conversations (ERC)~\cite{poria2019} for multi-turn dialogues. Early text-based methods used dictionaries, machine learning, or BERT. However, human emotion is inherently multimodal, involving speech, face, and body language. With the growth of multimedia, integrating visual, auditory, and textual information has become a focus.

Recent multimodal large language models (MLLMs), such as GPT-4V, Qwen-VL, and LLaVA, offer new possibilities. Pre-trained on image-text pairs, they process video and text end-to-end, directly capturing emotional cues without pre-extracted features~\cite{shou2025}.

To enhance MLLMs for sentiment analysis, several approaches have targeted unified representation learning. Rahman et al.~\cite{rahman2020} integrated multimodal cues into pretrained transformers, Guo et al.~\cite{guo2022} dynamically aligned unpaired inputs, and Hasan et al.~\cite{hasan2023} textualized multimodal information for better LLM utilization. Despite these advances, large labeled datasets and task-specific retraining remain costly. To address this, Yang et al.~\cite{yang2025} proposed MSE-Adapter, a lightweight plug-in for efficient multimodal handling, and Cai et al.~\cite{cai2025} designed hierarchical multi-layer fusion to jointly model multimodal features. Nevertheless, fixed fusion strategies and text-heavy reliance still limit adaptability to diverse inputs.

Moreover, MLLMs are highly sensitive to prompt design. Small changes in wording or demonstrations can lead to very different predictions. As shown in Figure 1, random demonstrations cause the model to output Neutral, a wrong result, while adaptively retrieved and contextually aligned demonstrations correctly identify Angry. Therefore, selecting adaptive multimodal demonstrations relevant to the query is crucial for eliciting MLLMs' sentiment reasoning under few-shot prompting.

\begin{figure}[!t]
    \centering
    \includegraphics[width=1\linewidth]{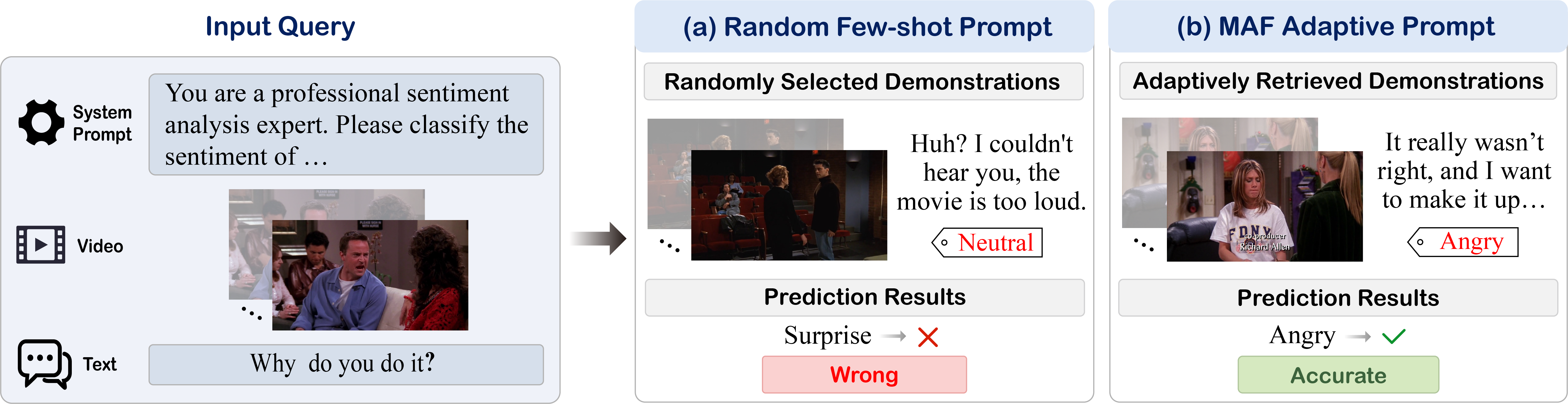}
    \caption{Illustration of sentiment analysis results from different prompts using ChatGLM3-6B. MAF adaptively retrieves multimodal demonstrations aligned with the query, yielding more accurate and consistent predictions.}
    \label{fig:process}
\end{figure}

To address these issues, we propose the \textbf{M}ultimodal \textbf{A}daptive \textbf{F}ew-shot Prompting (\textbf{MAF}) framework. It does not update the parameters of the deployed MLLM; instead, it combines retrieval-augmented prompting with a light-weight trainable coefficient generator to achieve dynamic demonstration retrieval and adaptive weight fusion.

The main contributions of this paper are as follows:
\begin{enumerate}
    \item We design a multi-feature demonstration retrieval mechanism that integrates facial (with lip motion for speaker localization), scene, and text features to precisely retrieve complementary multimodal demonstrations.
    \item We propose an adaptive coefficient generation method that trains a light-weight network to output optimal fusion weights per query for personalized similarity aggregation.
    \item We establish a majority-voting inference mechanism that generates multiple candidate outputs and votes to stabilize predictions, thereby improving robustness.
\end{enumerate}

\section{Related Work}

MSA and ERC have recently been advanced by MLLMs. Research on enhancing MLLMs for sentiment analysis falls into three categories: retrieval-augmented example selection, LLM-based multimodal sentiment analysis, and output stability enhancement.

\subsection{Retrieval-Augmented Example Selection}

Retrieval-Augmented Generation (RAG) has demonstrated strong capability in enhancing large language models. Originally developed for open-domain question answering, RAG retrieves contextually relevant passages and incorporates them into the model input, improving few-shot and zero-shot performance~\cite{karpukhin2020dense}. Subsequent works have proposed various optimizations, including latent retrieval for weakly supervised settings~\cite{lee2019latent} and end-to-end differentiable retrieval with improved query-aware similarity scoring~\cite{lewis2020rag}. These advances allow RAG to enhance LLM predictions without updating model parameters. Although most prior work focuses on textual retrieval, the underlying principle can be extended to multimodal scenarios, where visual, auditory, and textual features could jointly guide example selection. Building on this idea, MAF applies a RAG-style retrieval strategy for multimodal sentiment analysis, dynamically selecting the most informative demonstrations across modalities to improve prediction quality.

\subsection{Large Language Models in Multimodal Sentiment Analysis}

Large language models (LLMs) such as GPT-4 and LLaMA have demonstrated strong few-shot and zero-shot reasoning capabilities on textual inputs~\cite{brown2020,radford2023}. Their ability to process and generate natural language enables effective understanding of sentiment-related context, but they are inherently limited to purely textual information~\cite{yin2023survey}. Recent advances in multimodal large language models (MLLMs), together with multimodal prompting frameworks built on Qwen~\cite{bai2023qwen}, LLaMA2~\cite{touvron2023llama}, and ChatGLM3~\cite{glm2024chatglm}, enable the incorporation of visual cues, providing additional complementary information for sentiment understanding. Lightweight multimodal integration methods such as MAG-BERT~\cite{rahman2020} and MSE-Adapter~\cite{yang2025} provide lightweight mechanisms to extend existing LLMs for multimodal inputs without full model fine-tuning, offering a parameter-efficient complement to full MLLM training. Despite these advances, current approaches remain sensitive to prompt quality and modality importance, which can result in unstable predictions. To address these challenges, MAF dynamically retrieves and fuses multimodal demonstrations conditioned on the input query, performing adaptive weighted aggregation over facial, scene and text features to improve prediction stability and accuracy.

\subsection{Output Stability Enhancement}

MLLM outputs are inherently unstable due to probabilistic decoding~\cite{holtzman2020}. To mitigate this, self-consistency~\cite{wang2023} and structured prompting with LLM ensembling~\cite{gao2025} have been explored, improving robustness and reliability. In dataset construction, majority voting among multiple annotators has been shown to reduce label ambiguity and enhance annotation reliability~\cite{lian2023}, inspiring inference-time aggregation strategies. Motivated by these insights, MAF adopts majority voting over multiple sampled outputs to improve output stability in multimodal sentiment prediction.

\begin{figure}[!t]
    \centering
    \includegraphics[width=1\linewidth]{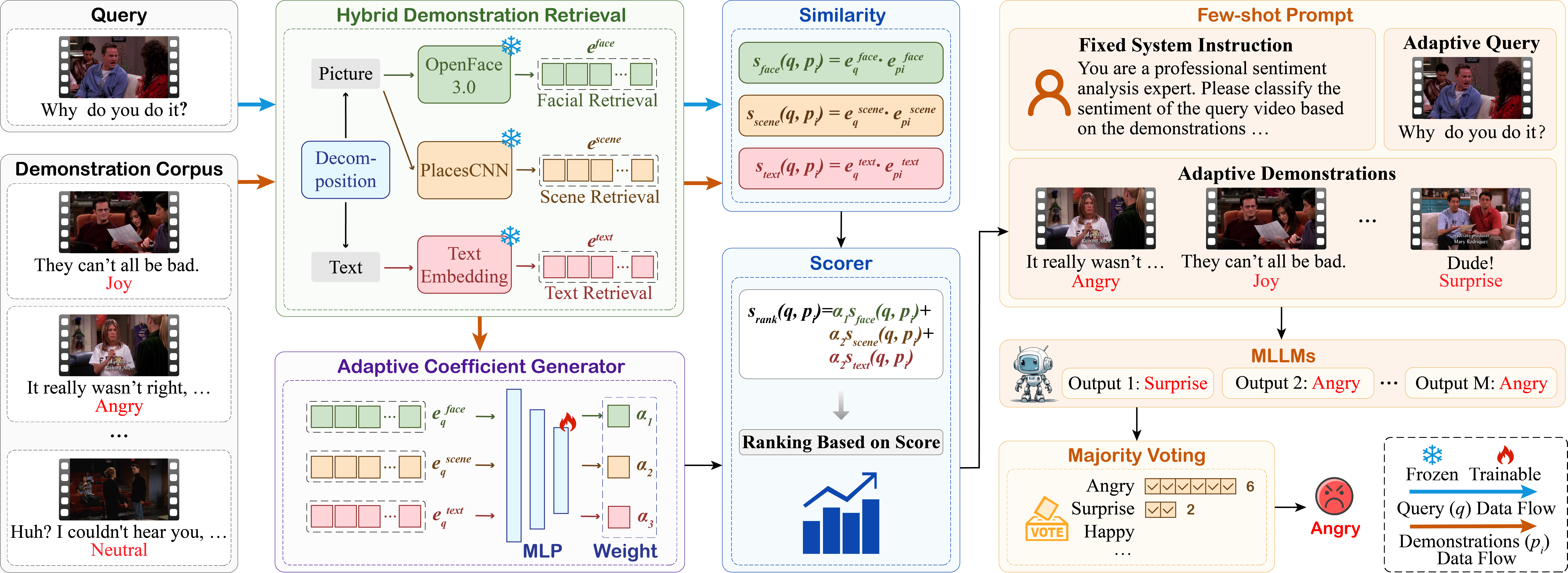}
    \caption{The overview of the MAF framework.}
    \label{fig:frame}
\end{figure}

\section{Method}

MAF first adaptively retrieves appropriate demonstrations from a multimodal corpus and directly embeds them into a prompt template. It then conducts majority voting on multiple candidate outputs generated by a deployed large language model to obtain the final sentiment analysis results. As shown in Figure~\ref{fig:frame}, MAF consists of three core components: the hybrid demonstration retrieval module, the adaptive coefficient generation module, and the majority voting module. Each of these modules will be elaborated on in detail in the following sections. In this study, besides the embedded demonstrations and query content, the prompt template employs fixed prompts to describe the general task.

\subsection{Prompt Templates and Demonstration Corpus}

\begin{figure}[!t]
    \centering
    \includegraphics[width=1.0\linewidth]{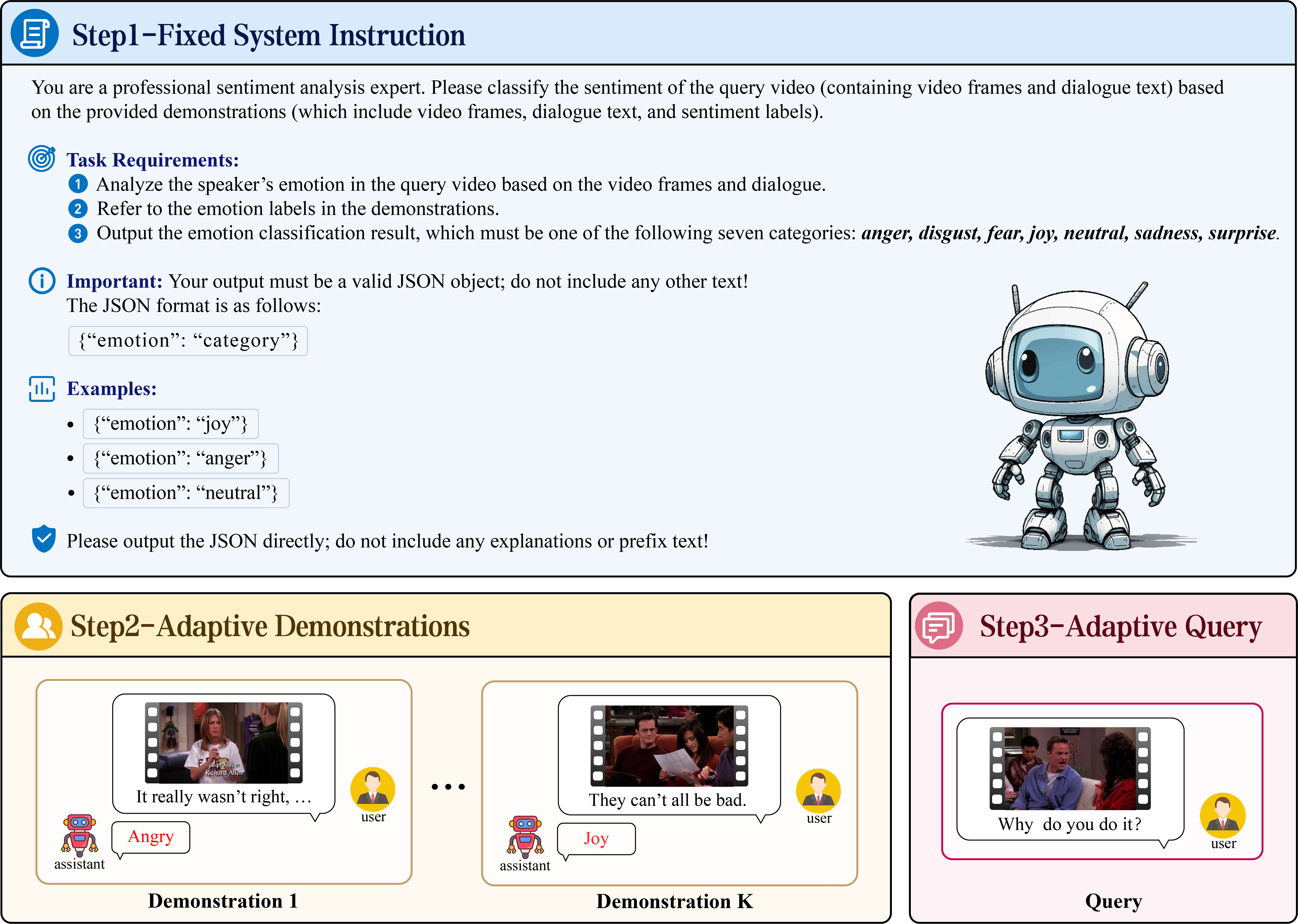}
    \caption{Illustration of the few-shot prompt used in our work.}
    \label{fig:prompt}
\end{figure}

MAF employs a fixed multi-turn dialogue structure. Instead of engineering task-specific prompt templates, we achieve personalization by adaptively retrieving and injecting relevant multimodal demonstrations for each query. As illustrated in Figure~\ref{fig:prompt}, the template comprises three fixed components: a system message defining the task and output format; demonstration slots dynamically populated with retrieved video-text-label pairs; and the query message containing the target video and text. During inference, the overall dialogue structure remains static; adaptability stems entirely from the retrieved demonstrations, preserving a concise and generalizable template.

For each dataset, we separately construct the demonstration corpus \(\mathcal{D} = \{ \langle d_i^{img}, d_i^{txt}, d_i^{label} \rangle \}_{i=1}^{N}\), where \(N=1000\) in our experiments, by randomly sampling instances from the training splits. To avoid data leakage, \(\mathcal{D}\) is disjoint from the validation set, test set, and the training samples for the adaptive coefficient generator in Section~\ref{sec:adaptive_generator}. In practical deployment, this corpus can be readily expanded with additional high-quality annotated data.

\subsection{Hybrid Demonstration Retrieval Module}

This module retrieves the most relevant multimodal demonstrations to populate the prompt template. Since demonstration quality and relevance directly govern sentiment analysis performance, constructing highly discriminative multimodal representations is critical. To this end, we propose a multi-feature hybrid retrieval module that ranks candidates using a weighted similarity of facial, scenes, and text features.

Specifically, given an input query \( q \) (comprising an image sequence \( q^{img} = \{ q_t^{img} \}_{t=1}^{T} \) and text \( q^{txt} \)), the module retrieves the \( K \) most similar multimodal demonstrations from \( \mathcal{D} \):

\begin{equation}
\{ \langle d_{k}^{img}, d_{k}^{txt}, d_{k}^{label} \rangle \}_{k=1}^{K} = f_{hybrid}(q, \mathcal{D})
\end{equation}

where \( \mathcal{D} = \{ \langle d_{i}^{img}, d_{i}^{txt}, d_{i}^{label} \rangle \}_{i=1}^{N} \) contains \( N \) demonstrations, and \( f_{hybrid}(\cdot) \) denotes the multi-feature similarity fusion function. 

For the query image sequence \( q^{img} \), we first extract three granularities of feature representations:

\textbf{Facial Features} \( e_q^{face} \): Extracted per frame \( q_t^{img} \) using OpenFace3.0. For multi-person scenes, we localize the active speaker by calculating the sum of Euclidean distance changes between upper and lower lip keypoints across consecutive frames. Temporal average pooling is then applied to obtain the video-level feature:
    \begin{equation}
    e_q^{face} = \frac{1}{T} \sum_{t=1}^{T} e_{q,t}^{face}
    \end{equation}

 \textbf{Scene Features} \( e_q^{scene} \): Extracted via PlaceCNN for each frame, followed by temporal average pooling:
    \begin{equation}
    e_q^{scene} = \frac{1}{T} \sum_{t=1}^{T} e_{q,t}^{scene}
    \end{equation}

\textbf{Text Features} \( e_q^{text} \): Extracted using an LLM text encoder. The input text is mapped into an embedding sequence \( E_q \), followed by max-pooling and normalization:
    \begin{equation}
    e_q^{text} = \operatorname{norm}(\operatorname{MaxPooling}(E_q))
    \end{equation}

Candidate demonstrations \( p_i = \langle d_i^{img}, d_i^{txt}, d_i^{label} \rangle \) in \( \mathcal{D} \) are processed identically to obtain \( e_{p_i}^{face} \), \( e_{p_i}^{scene} \), and \( e_{p_i}^{text} \). We then compute three similarity scores between \( q \) and each \( p_i \) via inner products:

\textbf{Facial Feature Similarity:}
    \begin{equation}
    s_{face}(q, p_i) = e_q^{face} \cdot e_{p_i}^{face}
    \end{equation}
    
\textbf{Scene Feature Similarity:}
    \begin{equation}
    s_{scene}(q, p_i) = e_q^{scene} \cdot e_{p_i}^{scene}
    \end{equation}
    
\textbf{Text Feature Similarity:}
    \begin{equation}
    s_{text}(q, p_i) = e_q^{text} \cdot e_{p_i}^{text}
    \end{equation}

The final retrieval score \( s_{rank}(q, p_i) \) is obtained through weighted fusion:
\begin{equation}
s_{rank}(q, p_i) = \alpha_1 s_{face}(q, p_i) + \alpha_2 s_{scene}(q, p_i) + \alpha_3 s_{text}(q, p_i)
\label{eq:rank}
\end{equation}
where \( \alpha_1, \alpha_2, \alpha_3 \) are non-negative fusion coefficients. In MAF, these coefficients are generated adaptively for each query by the coefficient generator described in Section~\ref{sec:adaptive_generator}. We rank all candidates by \( s_{rank} \) and select the top-\(K\) demonstrations to populate the prompt.

\subsection{Adaptive Coefficient Generator}
\label{sec:adaptive_generator}

Using fixed fusion coefficients \(\alpha_1,\alpha_2,\alpha_3\) in Eq.~\ref{eq:rank} is suboptimal for accommodating diverse sentiment expression patterns. To dynamically optimize this fusion strategy, we propose an adaptive coefficient generator—a lightweight neural network that predicts optimal fusion weights in real time based on the input query content. 

\subsubsection{Model Architecture}

The generator takes the concatenated multi-feature representation of query \(q\) as input. Specifically, \(e_q^{face} \in \mathbb{R}^{512}\), \(e_q^{scene} \in \mathbb{R}^{2048}\), and \(e_q^{text} \in \mathbb{R}^{1536}\) are concatenated into a joint vector:
\begin{equation}
    \mathbf{h}_q = [e_q^{face}; e_q^{scene}; e_q^{text}] \in \mathbb{R}^{4096}
\end{equation}

\(\mathbf{h}_q\) is then fed into a two‑layer MLP with a 256‑dimensional hidden layer and ReLU activation, followed by a sigmoid output layer that produces independent scores for each candidate weight combination:
\begin{equation}
\mathbf{p}_q = \sigma\bigl(\mathbf{W}_2 \cdot \operatorname{ReLU}(\mathbf{W}_1 \mathbf{h}_q + \mathbf{b}_1) + \mathbf{b}_2\bigr) \in [0,1]^{C}
\end{equation}
where \(C\) is the number of predefined weight combinations and \(\sigma(\cdot)\) is the element-wise sigmoid function. Each entry \(p_{q,c}\) indicates the predicted probability that the \(c\)-th combination is correct for query \(q\). During inference, the final fusion coefficients \([\alpha_1,\alpha_2,\alpha_3]\) are taken as the combination with the highest score.

\subsubsection{Training Objective}

We discretize the continuous weight space with a step size of \(0.1\) under the constraints \(\alpha_1+\alpha_2+\alpha_3=1\) and \(\alpha_i \ge 0\), yielding \(C = 66\) distinct triples. For each training sample \(x_j\) and each triple \(c\), we run the MAF framework and record correctness \(t_{j,c}=\mathbf{1}[\hat{y}_{j,c}=y_j^{\text{true}}]\). Note that multiple triples may be correct for the same sample, making this a multi‑label setting. The generator is trained with binary cross‑entropy loss:
\begin{equation}
\mathcal{L} = -\frac{1}{N_{\text{train}}}\sum_{j=1}^{N_{\text{train}}}\sum_{c=1}^{C} \bigl[ t_{j,c} \log p_{j,c} + (1-t_{j,c}) \log(1-p_{j,c}) \bigr]
\end{equation}
The model is optimized with Adam, learning rate \(10^{-3}\), and dropout \(p=0.2\) until the validation loss converges.

\subsubsection{Inference Phase}

During inference, the trained generator efficiently produces query‑specific fusion coefficients by selecting the weight combination with the highest predicted probability: \(\boldsymbol{\alpha}^* = \boldsymbol{\alpha}^{(c^*)}\) where \(c^* = \arg\max_c \mathbf{p}_q[c]\). These adaptive weights guide the hybrid retrieval module to select the most relevant demonstrations, seamlessly integrating into the end‑to‑end MAF framework.

\subsection{Majority Voting Module}

Multimodal LLMs typically employ probabilistic decoding, which yields semantic diversity but introduces potential instability in predictions. To enhance robustness, we design a majority voting module to aggregate multiple generated candidates into a reliable final prediction.

Given query \( q \) and \( K \) retrieved demonstrations, we formulate a complete prompt \( \mathcal{P}(q) \). The LLM generates \( M \) candidate sentiment labels via probabilistic sampling:
\begin{equation}
\{ \hat{y}_m \}_{m=1}^{M} = \operatorname{LLM}(\mathcal{P}(q)), \quad \hat{y}_m \in \mathcal{Y}
\end{equation}
where \( \mathcal{Y} \) is the label space. The module identifies the mode of these candidates as the final prediction:
\begin{equation}
y^* = \arg\max_{y \in \mathcal{Y}} \sum_{m=1}^{M} \mathbb{I}(\hat{y}_m = y)
\end{equation}
where \( \mathbb{I}(\cdot) \) is the indicator function. Ties are resolved randomly, which empirically suffices to mitigate the impact of outlier predictions. This training-free mechanism operates without additional parameters, finalizing the robust MAF framework. Moreover, thanks to parallel API calls, generating \(M\) candidates incurs nearly the same latency as a single deterministic call.

\section{Experiments}

\subsection{Experimental Settings}

\subsubsection{Datasets}

To comprehensively evaluate the effectiveness of the proposed MAF framework on multimodal sentiment analysis tasks, we conducted experiments on three public benchmark datasets: CMU-MOSEI~\cite{zadeh2018}, CH-SIMS v2.0~\cite{liu2022}, and MELD~\cite{poria2019}. These three datasets complement each other well in terms of language, dialogue format, and annotation types: CMU-MOSEI provides English monologue scenarios, CH-SIMS v2.0 provides diverse Chinese scenarios, and MELD provides English multi-speaker dialogue scenarios. Detailed statistics for each dataset are shown in Table~\ref{tab:datasets}.

\begin{table*}[!t]
\centering
\small
\caption{Characteristics of sentiment and emotion datasets. The table includes the dataset name, source, language, total number of samples, and label types.}
\label{tab:datasets}
\adjustbox{max width=\textwidth}{%
\begin{tabular}{c|c|c|c|c}
\toprule
\multicolumn{1}{c|}{Dataset} & \multicolumn{1}{c|}{Source} & \multicolumn{1}{c|}{Language} & \multicolumn{1}{c|}{Total Samples} & \multicolumn{1}{c}{Label Types} \\
\midrule
CMU-MOSEI & YouTube & English & 23,453 & [-3, +3] \\
CH-SIMS v2.0 & Movies, TV series and variety shows & Chinese & 4,361 & [-1, +1] \\
MELD & Friends & English & 13,707 & 7 emotion categories \\
\bottomrule
\end{tabular}%
}
\end{table*}

\subsubsection{Baselines}

We compared the performance of MAF against previous state-of-the-art models: TFN~\cite{zadeh2017}, LMF~\cite{liu2018}, MulT~\cite{tsai2019}, MAG-BERT~\cite{rahman2020}, Self-MM~\cite{yu2021}, AV-MC~\cite{liu2022}, UFEN-MTFN~\cite{cai2025}, TextMI~\cite{hasan2023}, CHFN~\cite{guo2022}, UniMSE~\cite{hu2022}, UniSAGPT2, UniSAT5, and UniSABART~\cite{li2023b}, MMGCN~\cite{hu2021}, MM-DFN~\cite{hu2022a}, EmoKEG~\cite{emokeg}, GA2MIF~\cite{li2023a}, MSE~\cite{yang2025}, and ATGFB~\cite{liu2026atgfb}.

\subsubsection{Metrics}
Due to differences in task settings, we adopt dataset-specific evaluation metrics. For CMU-MOSEI, we report MAE, Corr, Acc-7, Acc-2, and F1, where Acc-2 and F1 are computed under the non-negative/negative setting. For CH-SIMS v2.0, we report MAE, Corr, Acc-2, F1, and \(\mathrm{Acc2}_{\mathrm{weak}}\), where \(\mathrm{Acc2}_{\mathrm{weak}}\) evaluates weak sentiment samples in the range [-0.4, 0.4]. For MELD, we report Acc and WF1. Except for MAE, higher values indicate better performance.

\subsubsection{Implementation Details}

To validate the effectiveness of MAF, we conducted experiments on three representative LLM/MLLM backbones used in multimodal sentiment settings: Qwen-1.8B, LLaMA2-7B, and ChatGLM3-6B. The FAISS vector database was adopted, and search efficiency was optimized using FAISS's approximate nearest neighbor search, thereby improving search performance. We use concurrent generation for the voting module, so the wall-clock latency of multiple candidate generations remains close to that of a single generation in our implementation. Unless otherwise stated, we set \(K=3\) and \(M=8\), following the hyperparameter analysis in Section 4.2.  All experiments were run on NVIDIA GeForce RTX 4090. To ensure reproducibility, all experiments used a fixed seed.

\subsection{Experimental Results}

\subsubsection{Main Result}

Table~\ref{tab:mresult} compares MAF with representative multimodal sentiment analysis and emotion recognition baselines on CMU-MOSEI, CH-SIMS v2.0, and MELD. Overall, MAF consistently improves the corresponding MSE and ATGFB variants under the same backbone in most metrics, demonstrating that adaptive demonstration retrieval and voting-based inference are effective across different model families. On CMU-MOSEI, MAF-ChatGLM3-6B achieves the strongest Acc-2, F1, Acc-7, and Corr scores, reaching 90.23\%, 89.83\%, 59.58\%, and 0.805, respectively. Meanwhile, MAF-LLaMA2-7B obtains the lowest MAE of 0.475, indicating its advantage in regression-oriented sentiment intensity prediction. On CH-SIMS v2.0, MAF-ChatGLM3-6B obtains the best Acc-2 and F1 scores among the compared methods, with 86.89\% and 85.94\%, respectively, and also achieves the lowest MAE of 0.279. Although ATGFB-ChatGLM3-6B reports a higher \(\mathrm{Acc2}_{\mathrm{weak}}\) score, MAF maintains better overall performance across the major classification and regression metrics. On MELD, MAF-ChatGLM3-6B reaches 69.94\% Acc and 69.25\% WF1, outperforming previous ERC baselines and the corresponding MSE/ATGFB variants. These results show that MAF is not a backbone-specific improvement, but a general prompting framework that can enhance different MLLMs through query-aware demonstration selection and robust output aggregation.

\begin{table}[!t]
\centering
\small
\caption{Performance comparison of MAF against baseline methods. Within each backbone block, the best results are highlighted in \textbf{bold}.}
\label{tab:mresult}
\adjustbox{max width=\textwidth}{%
\begin{tabular}{c|ccccc|ccccc|cc}
\toprule
\multirow{2}{*}{Model} & \multicolumn{5}{c|}{MOSEI} & \multicolumn{5}{c|}{CH-SIMS v2.0} & \multicolumn{2}{c}{MELD} \\
\cmidrule(lr){2-6} \cmidrule(lr){7-11} \cmidrule(lr){12-13}
& Acc-2 & F1 & Acc-7 & MAE & Corr & Acc-2 & F1 & \(\mathrm{Acc2}_{\mathrm{weak}}\) & MAE & Corr & Acc & WF1 \\
\midrule
TFN           & 78.50 & 78.96 & 51.60 & 0.573 & 0.714 & 76.51 & 76.31 & 66.27 & 0.323 & 0.667 & 60.77 & 57.74 \\
LMF           & 80.54 & 80.94 & 51.59 & 0.576 & 0.717 & 77.05 & 77.02 & 69.34 & 0.343 & 0.638 & 61.15 & 58.30 \\
MulT          & 81.15 & 81.56 & 52.84 & 0.559 & 0.733 & 79.50 & 79.59 & 69.61 & 0.317 & 0.703 & -    & -     \\
MAG-BERT      & 82.51 & 82.77 & 50.41 & 0.583 & 0.741 & 79.79 & 79.78 & 71.87 & 0.334 & 0.691 & -    & -     \\
Self-MM       & 82.81 & 82.53 & 53.46 & 0.530 & 0.765 & 79.01 & 78.89 & 71.87 & 0.335 & 0.640 & -    & -     \\
AV-MC         & -     & -     & -     & -     & -     & 82.50 & 82.55 & 74.54 & 0.297 & 0.732 & -    & -     \\
UFEN-MTFN     & 84.80 & 84.90 & 54.50 & 0.535 & 0.760 & -    & -    & -    & -    & -    & -    & -     \\
TextMI        & 86.45 & 86.38 & 53.21 & 0.531 & 0.774 & -    & -    & -    & -    & -    & -    & -     \\
CHFN          & 83.70 & 83.90 & 54.30 & 0.525 & 0.778 & -    & -    & -    & -    & -    & -    & -     \\
UniMSE        & 85.86 & 85.79 & 54.39 & 0.523 & 0.773 & -    & -    & -    & -    & -    & 65.09 & 65.51 \\
UniSAGPT2     & 71.02 & -    & 41.36 & 0.838 & -    & -    & -    & -    & -    & -    & 48.12 & 31.26 \\
UniSAT5       & 84.22 & -    & 52.50 & 0.546 & -    & -    & -    & -    & -    & -    & 64.52 & 62.17 \\
UniSABART     & 84.93 & -    & 50.03 & 0.587 & -    & -    & -    & -    & -    & -    & 62.34 & 62.22 \\
MMGCN         & -    & -    & -    & -    & -    & -    & -    & -    & -    & -    & 60.42 & 58.31 \\
MM-DFN        & -    & -    & -    & -    & -    & -    & -    & -    & -    & -    & 62.49 & 59.46 \\
EmoKEG       & -    & -    & -    & -    & -    & -    & -    & -    & -    & -    & 66.44    & 65.69 \\
GA2MIF        & -    & -    & -    & -    & -    & -    & -    & -    & -    & -    & 61.65 & 58.94 \\ 
\midrule
MSE-Qwen-1.8B    & 84.12 & 83.45 & 52.02 & 0.558 & 0.725 & 80.44 & 80.24 & 73.09 & \textbf{0.311} & 0.678 & 62.18 & 59.87 \\
ATGFB-Qwen-1.8B    & 87.75 & 87.52 &  54.23 & 0.531 & 0.779 & 80.82 &  80.67 & 74.50 &  0.362 & 0.692 & 65.48 & 62.56 \\
\textbf{MAF-Qwen-1.8B} & \textbf{87.98} & \textbf{87.62} & \textbf{54.48} & \textbf{0.520} & \textbf{0.783} & \textbf{80.94} & \textbf{80.81} & \textbf{74.65} & 0.338 & \textbf{0.729} & \textbf{65.52} & \textbf{63.28} \\
\midrule
MSE-LLaMA2-7B    & 86.74 & 86.51 & 55.57 & 0.501 & 0.787 & 75.53 & 75.44 & 68.61 & 0.382 & 0.553 & 65.14 & 63.66 \\
ATGFB-LLaMA2-7B    & 88.63 & 87.84 & 57.76 & 0.486 & \textbf{0.798} & 77.97 & 77.74 & 71.59 & 0.379 & 0.556 &  66.68 & 63.55 \\
\textbf{MAF-LLaMA2-7B} & \textbf{88.76} & \textbf{87.91} & \textbf{58.12} & \textbf{0.475} & 0.795 & \textbf{80.53} & \textbf{80.65} & \textbf{73.90} & \textbf{0.326} & \textbf{0.694} & \textbf{68.81} & \textbf{65.76} \\
\midrule
MSE-ChatGLM3-6B  & 86.91 & 86.77 & 54.56 & 0.515 & 0.783 & 83.77 & 83.76 & 75.24 & 0.296 & 0.720 & 66.23 & 65.13 \\
ATGFB-ChatGLM3-6B  & 89.83 & 89.49 & 58.04 & 0.529 & 0.803 & 86.85 & 85.75 & \textbf{79.43} &  0.284 & 0.736 & 69.91 & 68.93 \\
\textbf{MAF-ChatGLM3-6B} & \textbf{90.23} & \textbf{89.83} & \textbf{59.58} & \textbf{0.512} & \textbf{0.805} & \textbf{86.89} & \textbf{85.94} & 77.14 & \textbf{0.279} & \textbf{0.737} & \textbf{69.94} & \textbf{69.25} \\
\bottomrule
\end{tabular}%
}
\end{table}

\begin{table*}[!t]
\centering
\caption{The ablation experiment results on MOSEI, CH-SIMS v2.0, and MELD.}
\label{tab:ablation}
\adjustbox{max width=\textwidth}{%
\begin{tabular}{c|ccccc|ccccc|cc}
\toprule
\multirow{2}{*}{Model} 
& \multicolumn{5}{c|}{MOSEI} 
& \multicolumn{5}{c|}{CH-SIMS v2.0} 
& \multicolumn{2}{c}{MELD} \\
\cmidrule(lr){2-6} \cmidrule(lr){7-11} \cmidrule(lr){12-13}
& Acc-2 & F1 & Acc-7 & MAE & Corr 
& Acc-2 & F1 & \(\mathrm{Acc2}_{\mathrm{weak}}\) & MAE & Corr 
& Acc & WF1 \\
\midrule
w/o RAG                & 88.94 & 88.03 & 57.45 & 0.516 & 0.787 & 83.97 & 84.12 & 75.36 & 0.296 & 0.719 & 69.72 & 67.55 \\
w/o adaptive retrieval & 87.48 & 86.62 & 56.12 & 0.524 & 0.775 & 82.43 & 83.56 & 75.06 & 0.313 & 0.706 & 68.41 & 65.96 \\
w/o adaptive weighting & 88.71 & 87.96 & 56.68 & 0.521 & 0.781 & 82.48 & 83.68 & 74.93 & 0.318 & 0.715 & 69.17 & 67.29 \\
w/o voting             & 87.30 & 86.11 & 55.37 & 0.534 & 0.771 & 81.72 & 81.89 & 73.55 & 0.330 & 0.697 & 67.65 & 65.50 \\
\midrule
MAF                    & \textbf{90.23} & \textbf{89.83} & \textbf{59.58} & \textbf{0.512} & \textbf{0.805} & \textbf{86.89} & \textbf{85.94} & \textbf{77.14} & \textbf{0.279} & \textbf{0.737} & \textbf{69.94} & \textbf{69.25} \\
\bottomrule
\end{tabular}%
}
\end{table*}

\subsubsection{Ablation Study}

To evaluate the contribution of each component, we conduct ablation experiments on CMU-MOSEI, CH-SIMS v2.0, and MELD using ChatGLM3-6B as the backbone. As shown in Table~\ref{tab:ablation}, the complete MAF framework achieves the best overall performance on all three datasets, confirming that the proposed retrieval, weighting, and voting modules are complementary.

Removing the RAG-based demonstration mechanism causes a noticeable drop on CH-SIMS v2.0 where Acc-2 falls from 86.89\% to 83.97\% but only a marginal change on MELD where Acc goes from 69.94\% to 69.72\%. More strikingly, removing adaptive retrieval leads to even larger degradation across all datasets. For example, Acc-2 on MOSEI drops from 90.23\% to 87.48\%, and on CH-SIMS v2.0 from 86.89\% to 82.43\%. This reveals that the accuracy of demonstrations is crucial, as erroneous ones may mislead the model. Removing adaptive weighting also consistently hurts performance, especially on MAE for CH-SIMS v2.0 from 0.279 to 0.318 and for MOSEI from 0.512 to 0.521, showing that fixed fusion weights cannot handle diverse sentiment cues. Removing the voting module causes a clear drop on MOSEI and CH-SIMS v2.0 but a smaller effect on MELD, where Acc declines from 69.94\% to 67.65\%, reflecting that large models are prone to fluctuation, making majority voting especially beneficial.

\begin{figure}[!t]
    \centering
    \includegraphics[width=0.9\linewidth]{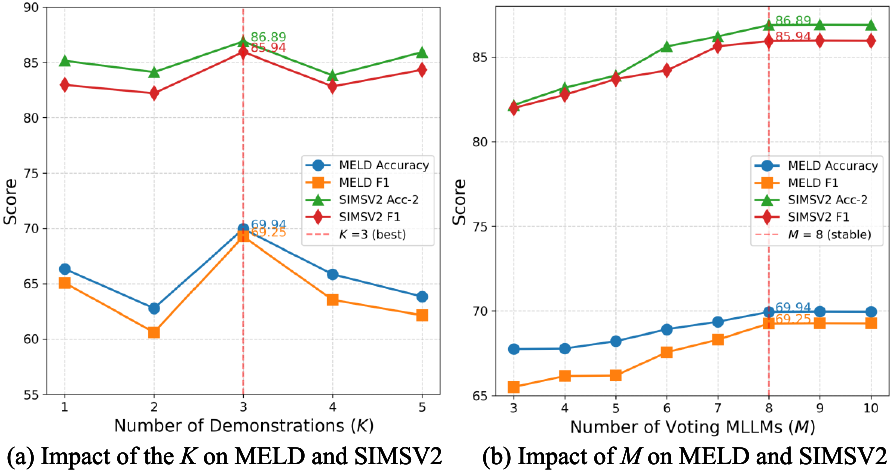}
    \caption{Impact of the number of retrieved demonstrations ($K$) and the number of voting candidates ($M$) on MELD and CH-SIMS v2.0. The best trade-off is achieved with $K=3$ and a stable voting setting around $M=8$.}
    \label{fig:top_k}
\end{figure}

\subsubsection{Impact of top-\(K\) and Voting Number}

We further analyze two key hyperparameters in MAF: the number of retrieved demonstrations $K$ and the number of sampled candidate outputs $M$. As shown in Figure~\ref{fig:top_k}, the left subfigure varies $K$ from 1 to 5. The performance on MELD and CH-SIMS v2.0 reaches its best value when $K=3$, indicating that three demonstrations provide sufficient contextual guidance without introducing excessive irrelevant examples. When $K$ is smaller, the prompt lacks enough task-specific references; when $K$ is larger, less relevant demonstrations may dilute the useful sentiment cues.

The right subfigure studies the effect of $M$ in the majority voting module. Increasing $M$ generally improves performance at the beginning, because multiple sampled outputs reduce the influence of occasional unstable predictions. The curves become stable when $M$ is around 8, after which the additional gain becomes marginal. Therefore, we set $K=3$ and use the voting setting around $M=8$ in our experiments, balancing prediction accuracy, robustness, and inference cost.

\section{Conclusion}
This paper proposes the MAF framework to address prompt sensitivity, limited demonstration retrieval, and unstable outputs in MLLM-based sentiment analysis. MAF introduces three key components: a multi-feature hybrid retrieval mechanism that integrates facial, scene, and text features with lip-motion-based speaker localization; an adaptive coefficient generator that dynamically assigns query-specific fusion weights; and a majority-voting inference strategy that improves prediction stability. Experiments on three public datasets show that MAF consistently improves the corresponding backbone variants and achieves competitive or superior performance against strong baselines. Ablation results further confirm that retrieval quality, adaptive weighting, and voting-based aggregation provide complementary gains.

\bibliographystyle{splncs04}
\bibliography{ref}

@inproceedings{zadeh2018,
  author    = {AmirAli Bagher Zadeh and Paul Pu Liang and Soujanya Poria and Erik Cambria and Louis-Philippe Morency},
  title     = {Multimodal Language Analysis in the Wild: {CMU-MOSEI} Dataset and Interpretable Dynamic Fusion Graph},
  booktitle = {Proceedings of the 56th Annual Meeting of the Association for Computational Linguistics (Volume 1: Long Papers)},
  pages     = {2236--2246},
  year      = {2018},
  address   = {Melbourne, Australia},
  publisher = {Association for Computational Linguistics},
  doi       = {10.18653/v1/P18-1208}
}

@inproceedings{poria2019,
  author    = {Soujanya Poria and Devamanyu Hazarika and Navonil Majumder and Gautam Naik and Erik Cambria and Rada Mihalcea},
  title     = {{MELD}: A Multimodal Multi-Party Dataset for Emotion Recognition in Conversations},
  booktitle = {Proceedings of the 57th Annual Meeting of the Association for Computational Linguistics},
  pages     = {527--536},
  year      = {2019},
  address   = {Florence, Italy},
  publisher = {Association for Computational Linguistics},
  doi       = {10.18653/v1/P19-1050}
}

@article{shou2025,
  author    = {Yuntao Shou and Tao Meng and Wei Ai and Keqin Li},
  title     = {Multimodal Large Language Models Meet Multimodal Emotion Recognition and Reasoning: A Survey},
  journal   = {arXiv},
  volume    = {2509.24322},
  year      = {2025},
  doi       = {10.48550/arXiv.2509.24322},
  note      = {arXiv preprint arXiv:2509.24322}
}

@article{cai2025,
  author    = {Yujian Cai and Xingguang Li and Yingyu Zhang and Jinsong Li and Fazheng Zhu and Lin Rao},
  title     = {Multimodal sentiment analysis based on multi-layer feature fusion and multi-task learning},
  journal   = {Scientific Reports},
  volume    = {15},
  pages     = {2126},
  year      = {2025},
  doi       = {10.1038/s41598-025-85859-6}
}

@inproceedings{rahman2020,
  author    = {Wasifur Rahman and Md Kamrul Hasan and Sangwu Lee and AmirAli Bagher Zadeh and Chengfeng Mao and Louis-Philippe Morency and Ehsan Hoque},
  title     = {Integrating Multimodal Information in Large Pretrained Transformers},
  booktitle = {Proceedings of the 58th Annual Meeting of the Association for Computational Linguistics},
  pages     = {2359--2369},
  year      = {2020},
  address   = {Online},
  publisher = {Association for Computational Linguistics},
  doi       = {10.18653/v1/2020.acl-main.214}
}

@inproceedings{guo2022,
  author    = {Jiwei Guo and Jiajia Tang and Weichen Dai and Yu Ding and Wanzeng Kong},
  title     = {Dynamically Adjust Word Representations Using Unaligned Multimodal Information},
  booktitle = {Proceedings of the 30th ACM International Conference on Multimedia},
  pages     = {3394–3402},
  year      = {2022},
  publisher = {Association for Computing Machinery},
  address   = {New York, NY, USA},
  doi       = {10.1145/3503161.3548137}
}

@article{hasan2023,
  author  = {Md Kamrul Hasan and Md Saiful Islam and Sangwu Lee and Wasifur Rahman and Iftekhar Naim and Mohammed Ibrahim Khan and Ehsan Hoque},
  title   = {{TextMI}: Textualize Multimodal Information for Integrating Non-verbal Cues in Pre-trained Language Models},
  journal = {arXiv},
  volume  = {abs/2303.15430},
  year    = {2023}
}

@inproceedings{yang2025,
  author    = {Yang Yang and Xunde Dong and Yupeng Qiang},
  title     = {{MSE-Adapter}: A Lightweight Plugin Endowing LLMs with the Capability to Perform Multimodal Sentiment Analysis and Emotion Recognition},
  booktitle = {Proceedings of the AAAI Conference on Artificial Intelligence},
  volume    = {39},
  number    = {24},
  pages     = {25642--25650},
  year      = {2025},
  publisher = {Association for the Advancement of Artificial Intelligence},
  doi       = {10.1609/aaai.v39i24.34755}
}

@inproceedings{karpukhin2020dense,
  author    = {Vladimir Karpukhin and Barlas Oguz and Sewon Min and Patrick Lewis and Ledell Wu and Sergey Edunov and Danqi Chen and Wen-tau Yih},
  title     = {Dense Passage Retrieval for Open-Domain Question Answering},
  booktitle = {Proceedings of the 2020 Conference on Empirical Methods in Natural Language Processing (EMNLP)},
  pages     = {6769--6781},
  year      = {2020},
  address   = {Online},
  publisher = {Association for Computational Linguistics},
  doi       = {10.18653/v1/2020.emnlp-main.550}
}

@inproceedings{lee2019latent,
  author    = {Kenton Lee and Ming-Wei Chang and Kristina Toutanova},
  title     = {Latent Retrieval for Weakly Supervised Open Domain Question Answering},
  booktitle = {Proceedings of the 57th Annual Meeting of the Association for Computational Linguistics},
  pages     = {6086--6096},
  year      = {2019},
  publisher = {Association for Computational Linguistics},
  doi       = {10.18653/v1/P19-1612}
}

@inproceedings{lewis2020rag,
  author    = {Patrick Lewis and Ethan Perez and Aleksandra Piktus and Fabio Petroni and Vladimir Karpukhin and Naman Goyal and Heinrich Küttler and Mike Lewis and Wen-tau Yih and Tim Rocktäschel and Sebastian Riedel and Douwe Kiela},
  title     = {Retrieval-Augmented Generation for Knowledge-Intensive {NLP} Tasks},
  booktitle = {Advances in Neural Information Processing Systems},
  volume    = {33},
  pages     = {9459--9474},
  year      = {2020},
  publisher = {Curran Associates, Inc.}
}

@inproceedings{brown2020,
  author    = {Tom B. Brown and Benjamin Mann and Nick Ryder and Melanie Subbiah and Jared D. Kaplan and Prafulla Dhariwal and Arvind Neelakantan and Pranav Shyam and Girish Sastry and Amanda Askell and Sandhini Agarwal and Ariel Herbert-Voss and Gretchen Krueger and Tom Henighan and Rewon Child and Aditya Ramesh and Daniel M. Ziegler and Jeffrey Wu and Clemens Winter and Christopher Hesse and Mateusz Litwin and Scott Gray and Benjamin Chess and Jack Clark and Christopher Berner and Sam McCandlish and Alec Radford and Ilya Sutskever and Dario Amodei},
  title     = {Language Models are Few-Shot Learners},
  booktitle = {Advances in Neural Information Processing Systems},
  volume    = {33},
  pages     = {1877--1901},
  year      = {2020},
  publisher = {Curran Associates, Inc.}
}

@article{radford2023,
  title         = {GPT-4 Technical Report},
  author        = {{OpenAI}},
  journal       = {arXiv preprint},
  volume        = {arXiv:2303.08774},
  year          = {2023},
  doi           = {10.48550/arXiv.2303.08774}
}

@article{bai2023qwen,
    title     = {{Qwen-VL}: A Versatile Vision-Language Model for Understanding, Localization, Text Reading, and Beyond},
    author    = {Jinze Bai and Shuai Bai and Shusheng Yang and Shijie Wang and Sinan Tan and Peng Wang and Junyang Lin and Chang Zhou and Jingren Zhou},
    journal   = {arXiv},
    volume    = {arXiv:2308.12966},
    year      = {2023},
    doi       = {10.48550/arXiv.2308.12966}
}

@article{yin2023survey,
  author    = {Shukang Yin and Chaoyou Fu and Sirui Zhao and Ke Li and Xing Sun and Tong Xu and Enhong Chen},
  title     = {A Survey on Multimodal Large Language Models},
  journal   = {National Science Review},
  volume    = {11},
  number    = {12},
  pages     = {nwae403},
  year      = {2024},
  doi       = {10.1093/nsr/nwae403}
}

@inproceedings{holtzman2020,
  author    = {Ari Holtzman and Jan Buys and Li Du and Maxwell Forbes and Yejin Choi},
  title     = {The Curious Case of Neural Text Degeneration},
  booktitle = {International Conference on Learning Representations (ICLR)},
  year      = {2020}
}

@inproceedings{wang2023,
  author    = {Xuezhi Wang and Jason Wei and Dale Schuurmans and Quoc V Le and Ed H. Chi and Sharan Narang and Aakanksha Chowdhery and Denny Zhou},
  title     = {Self-Consistency Improves Chain of Thought Reasoning in Language Models},
  booktitle = {International Conference on Learning Representations (ICLR)},
  year      = {2023}
}

@inproceedings{gao2025,
  author = {Zhiqiang Gao and Shihao Gao and Zixing Zhang and Yihao Guo and Hongyu Chen and Jing Han},
  title     = {Structured Prompting and {LLM} Ensembling for Multimodal Conversational Aspect-based Sentiment Analysis},
  booktitle = {Proceedings of the 33rd ACM International Conference on Multimedia},
  pages = {14107–14113},
  year      = {2025},
  address   = {Dublin, Ireland},
  publisher = {Association for Computing Machinery},
  series = {MM '25},
  doi       = {10.1145/3746027.3762070}
}

@article{lian2023,
    title     = {Explainable Multimodal Emotion Reasoning},
    author    = {Zheng Lian and Haiyang Sun and Licai Sun and Hao Gu and Zhuofan Wen and Siyuan Zhang and Shun Chen and Mingyu Xu and Ke Xu and Kang Chen and Lan Chen and Shan Liang and Ya Li and Jiangyan Yi and Bin Liu and Jianhua Tao},
    journal   = {arXiv},
    volume    = {arXiv:2306.15401},
    year      = {2023},
    doi       = {10.48550/arXiv.2306.15401}
}

@inproceedings{zadeh2017,
  author    = {Amir Zadeh and Minghai Chen and Soujanya Poria and Erik Cambria and Louis-Philippe Morency},
  title     = {Tensor Fusion Network for Multimodal Sentiment Analysis},
  booktitle = {Proceedings of the 2017 Conference on Empirical Methods in Natural Language Processing (EMNLP)},
  pages     = {1103--1114},
  year      = {2017},
  address   = {Copenhagen, Denmark},
  publisher = {Association for Computational Linguistics},
  doi       = {10.18653/v1/D17-1115}
}

@inproceedings{liu2018,
  author    = {Zhun Liu and Ying Shen and Varun Bharadhwaj Lakshminarasimhan and Paul Pu Liang and Amir Zadeh and Louis-Philippe Morency},
  title     = {Efficient Low-rank Multimodal Fusion With Modality-Specific Factors},
  booktitle = {Proceedings of the 56th Annual Meeting of the Association for Computational Linguistics (Volume 1: Long Papers)},
  pages     = {2247--2256},
  year      = {2018},
  address   = {Melbourne, Australia},
  publisher = {Association for Computational Linguistics},
  doi       = {10.18653/v1/P18-1209}
}

@inproceedings{yu2021,
  author    = {Wenmeng Yu and Hua Xu and Ziqi Yuan and Jiele Wu},
  title     = {Learning Modality-Specific Representations with Self-Supervised Multi-Task Learning for Multimodal Sentiment Analysis},
  booktitle = {Proceedings of the AAAI Conference on Artificial Intelligence},
  volume    = {35},
  number    = {16},
  pages     = {10790--10797},
  year      = {2021},
  doi       = {10.1609/aaai.v35i12.17289}
}

@inproceedings{hu2022,
  author    = {Guimin Hu and Ting-En Lin and Yi Zhao and Guangming Lu and Yuzhuo Wu and Yongbin Li},
  title     = {{UniMSE}: Towards Unified Multimodal Sentiment Analysis and Emotion Recognition},
  booktitle = {Proceedings of the 2022 Conference on Empirical Methods in Natural Language Processing (EMNLP)},
  pages     = {7837--7851},
  year      = {2022},
  address   = {Abu Dhabi, United Arab Emirates},
  publisher = {Association for Computational Linguistics},
  doi       = {10.18653/v1/2022.emnlp-main.534}
}

@inproceedings{li2023b,
  author    = {Zaijing Li and Ting-En Lin and Yuchuan Wu and Meng Liu and Fengxiao Tang and Ming Zhao and Yongbin Li},
  title     = {{UniSA}: Unified Generative Framework for Sentiment Analysis},
  booktitle = {Proceedings of the 31st ACM International Conference on Multimedia (MM)},
  pages     = {6132--6142},
  year      = {2023},
  publisher = {Association for Computing Machinery},
  address   = {New York, NY, USA},
  series    = {MM '23}
}

@inproceedings{hu2021,
  author    = {Jingwen Hu and Yuchen Liu and Jinming Zhao and Qin Jin},
  title     = {{MMGCN}: Multimodal Fusion via Deep Graph Convolution Network for Emotion Recognition in Conversation},
  booktitle = {Proceedings of the 59th Annual Meeting of the Association for Computational Linguistics and the 11th International Joint Conference on Natural Language Processing (Volume 1: Long Papers)},
  pages     = {5666--5675},
  year      = {2021},
  address   = {Online},
  publisher = {Association for Computational Linguistics},
  doi       = {10.18653/v1/2021.acl-long.440}
}

@inproceedings{hu2022a,
  author    = {Dou Hu and Xiaolong Hou and Lingwei Wei and Lian-Xin Jiang and Yang Mo},
  title     = {{MM-DFN}: Multimodal Dynamic Fusion Network for Emotion Recognition in Conversations},
  booktitle = {ICASSP 2022 - 2022 IEEE International Conference on Acoustics, Speech and Signal Processing (ICASSP)},
  pages     = {7037--7041},
  year      = {2022},
  doi={10.1109/ICASSP43922.2022.9747397}}

@article{li2023a,
  author    = {Jiang Li and Xiaoping Wang and Guoqing Lv and Zhigang Zeng},
  title     = {{GA2MIF}: Graph and Attention Based Two-Stage Multiplexed Information Fusion for Conversational Emotion Detection},
  journal   = {IEEE Transactions on Affective Computing},
  volume    = {15},
  number    = {1},
  pages     = {130--143},
  year      = {2024},
  publisher = {IEEE}
}

@inproceedings{tsai2019,
  author    = {Yao-Hung Hubert Tsai and Shaojie Bai and Paul Pu Liang and J. Zico Kolter and Louis-Philippe Morency and Ruslan Salakhutdinov},
  title     = {Multimodal Transformer for Unaligned Multimodal Language Sequences},
  booktitle = {Proceedings of the 57th Annual Meeting of the Association for Computational Linguistics (ACL)},
  pages     = {6558--6569},
  year      = {2019},
  address   = {Florence, Italy},
  publisher = {Association for Computational Linguistics},
  doi       = {10.18653/v1/P19-1656}
}

@article{emokeg,
  title     = {{EmoKEG}: Knowledge-enhanced heterogeneous graph for emotion recognition in conversation},
  author    = {Lili Guo and Yanan Cui and Yikang Song and Haiwei Hou and Shifei Ding},
  journal   = {Pattern Recognition},
  volume    = {179},
  pages     = {113500},
  year      = {2026},
  publisher = {Elsevier},
  doi       = {10.1016/j.patcog.2026.113500}
}

@inproceedings{liu2026atgfb,
  title     = {{ATGFB-MFF}: Adaptive Text-Guided Fiber Bundle Feature Fusion with LLMs for Multimodal Sentiment Analysis and Emotion Recognition in Conversations},
  author    = {Zhaowei Liu and Sheng Liu and Weiqing Yan and Peng Song and Yongchao Song and Rufei Gao},
  booktitle={Proceedings of the ACM Web Conference 2026},
  pages={7442--7453},
  year={2026},
  publisher = {Association for Computing Machinery},
  address = {New York, NY, USA},
  doi = {10.1145/3774904.3792577},
  series = {WWW '26}
}

@article{touvron2023llama,
  title         = {Llama 2: Open Foundation and Fine-Tuned Chat Models},
  author        = {Hugo Touvron and Louis Martin and Kevin Stone and Peter Albert and Amjad Almahairi and Yasmine Babaei and Nikolay Bashlykov and Soumya Batra and Prajjwal Bhargava and Shruti Bhosale and Dan Bikel and Lukas Blecher and Cristian Canton Ferrer and Moya Chen and Guillem Cucurull and David Esiobu and Jude Fernandes and Jeremy Fu and Wenyin Fu and Brian Fuller and Cynthia Gao and Vedanuj Goswami and Naman Goyal and Anthony Hartshorn and Saghar Hosseini and Rui Hou and Hakan Inan and Marcin Kardas and Viktor Kerkez and Madian Khabsa and Isabel Kloumann and Artem Korenev and Punit Singh Koura and Marie-Anne Lachaux and Thibaut Lavril and Jenya Lee and Diana Liskovich and Yinghai Lu and Yuning Mao and Xavier Martinet and Todor Mihaylov and Pushkar Mishra and Igor Molybog and Yixin Nie and Andrew Poulton and Jeremy Reizenstein and Rashi Rungta and Kalyan Saladi and Alan Schelten and Ruan Silva and Eric Michael Smith and Ranjan Subramanian and Xiaoqing Ellen Tan and Binh Tang and Ross Taylor and Adina Williams and Jian Xiang Kuan and Puxin Xu and Zheng Yan and Iliyan Zarov and Yuchen Zhang and Angela Fan and Melanie Kambadur and Sharan Narang and Aurelien Rodriguez and Robert Stojnic and Sergey Edunov and Thomas Scialom},
  journal       = {arXiv},
  volume        = {arXiv:2307.09288},
  year          = {2023},
  doi           = {10.48550/arXiv.2307.09288}
}

@article{glm2024chatglm,
  title={ChatGLM: A Family of Large Language Models from {GLM-130B} to {GLM-4} All Tools}, 
  author={Team GLM},
  journal       = {arXiv},
  volume        = {arXiv:2406.12793},
  year          = {2024},
  doi           = {10.48550/arXiv.2406.12793}
}

@inproceedings{liu2022,
  author       = {Yihe Liu and Ziqi Yuan and Huisheng Mao and Zhiyun Liang and Wanqiuyue Yang and Yuanzhe Qiu and Tie Cheng and Xiaoteng Li and Hua Xu and Kai Gao},
  title        = {Make Acoustic and Visual Cues Matter: {CH-SIMS} v2.0 Dataset and {AV-Mixup} Consistent Module},
  booktitle    = {Proceedings of the 2022 International Conference on Multimodal Interaction (ICMI '22)},
  year         = {2022},
  pages        = {247--258},
  address   = {Bengaluru, India},
  publisher    = {ACM},
  doi          = {10.1145/3536221.3556630}
}

\end{document}